\documentclass[%
reprint,
superscriptaddress,
showkeys,
showpacs,preprintnumbers,
 amsmath,amssymb,
 aps,
]{revtex4-1}

\usepackage{graphicx}
\usepackage{dcolumn}
\usepackage{bm}
\usepackage{epstopdf}

\begin{document}


\title{Elliptic Davydov solitons in $\alpha$-helix protein chain with exciton-exciton and exciton-phonon couplings}

\author{Nkeh Oma Nfor}
\altaffiliation{Corresponding Author, Email: omnkeh@gmail.com}

\affiliation{Department of Physics, Higher Teacher Training College Bambili, The University of Bamenda, P. O. Box 39, Bambili-Cameroon.}

\author{Michael Nana Jipdi}

\affiliation{Department of Physics, Higher Teacher Training College Bambili, The University of Bamenda, P. O. Box 39, Bambili-Cameroon.}

\date{\today}
%
\begin{abstract}
We consider the Davydov model of $\alpha$-helix protein chain with both exciton-exciton and exciton-phonon couplings and investigate on the evolution of elliptic solitons. In the discrete regime of the adiabatic limit, we analytically and numerically show that modulational instability induces the self-localization of energy in the $\alpha$-helix protein chain. By incorporating the continuous limit approximations, various nonlinear periodic modes are traced; strongly suggesting that the energy of the ATP hydrolysis is locally distributed over the $\alpha$-helix protein chain. It is generally found that the exciton-exciton coupling induces the inhomogeneity in the protein chain, which greatly enhances energy localization that is physically manifested as nonlinear periodic modes. Results of numerical simulations clearly depicts the evolution of these nonlinear periodic modes in the highly discrete, nonlinear, and coupled system that governs the dynamics of $\alpha$-helix protein chain.
\end{abstract}
%
\maketitle

\section{Introduction}
The solitary wave concept as a localized and dynamic entity in nonlinear media plays a crucial role in modern science \cite{Peyrard2006}. This concept has greatly permeated the biophysical field because the first identification of biomedical soliton was made in $\alpha$-helix protein chains \cite{Davydov1973}. The Davydov soliton results from a nonlinear coupling between the vibrational excitation and a deformation in the protein chain, triggered by the amide-I vibrational energy \cite{Scott1992,Davydov1985,Christiansen1990}. This vibrational energy emanates from the hydrolysis of adenosine triphosphate (ATP). In fact, an ATP molecule binds to a specific site on the protein, reacts with water, and under normal physiological conditions releases the appropriate amount of energy for the metabolic activities of the cell. Concretely, Davydov mainly assumed that the amide-I molecules absorbed the energy released by the ATP and channelled it along the protein chain via quantum robust self-trapping excitations known as solitons. In this study, we mainly explore the quantum mechanical approach in dealing with the amide-I vibration, while classical mechanics is used in the lattice dynamics.

Despite the remarkable successes recorded by the Davydov model, there still exists some challenges like the exciton-phonon bound states constraint to travel only at subsonic velocity \cite{Scoot1982}, and the thermal instability of Davydov soliton at physiological temperatures \cite{Cottingham1989}. The issue related to the velocity has been appropriately addressed by the correction of the linear hydrogen bond potential which link peptides. For instance, the energy of hydrogen bond that links peptide groups was modelled by the cubic nonlinear potential \cite{Davydov1985,Davydov1983,Davydov1984}, with such modification giving an opportunity for the introduction of supersonic soliton velocity in the continuum limit approximation. To address the latter case, numerical analysis by Forner \cite{Forner1991} predicted the stability of Davydov soliton at physiological temperatures provided the nonlinear coupling constant is increased. Consequently this study seeks to explore additional mechanism for energy localization in the $\alpha$-helix protein chain, probably triggered by increase in nonlinearity. It should be noted that in the original model proposed by Davydov, anharmonicity is introduced in the protein chain only by the nonlinear coupling between excitons and the acoustic phonon field which describes the harmonic oscillations of the molecules along the chain. However, the effects of two exciton states in both the discrete and continuum levels in $\alpha$-helix protein chain greatly suggests that such dynamics induces additional nonlinearity in the chain \cite{Latha2012, Merlin2013}. The relatively large electric dipoles of the amide-I excitons further complement the idea of incorporating the exciton-exciton interaction in the protein chain \cite{Xiao1997,Nji2018}. It is important to note that in Ref. \cite{Xiao1997}, the exciton-exciton interaction is the driving force behind energy localization via the process of modulational instability in the $\alpha$-helix protein chain. The self-trapped amide-I vibrational energy may propagate along the chain as a result of the hydrolysis of ATP molecules.

A good number of research activities has been focused on identifying spatially localized soliton-type excitons in the $\alpha$-helix protein chain \cite{Davydov1973,Scott1992,Davydov1985,Christiansen1990}, with little focus on spatial periodic excitons. Such spatial periodic wave trains may be considered as a superposition of individual localized exciton modes. From experimental and theoretical standpoints, nonlinear periodic wavetrains are ubiquitous in nonlinear optics \cite{Rodrigues2018,Defi2015,Amrani2011,Haboucha2008}, neural networks \cite{Villacorta2013,Nfor2018}, liquid films \cite{Noumana2018}, among a plethora of nonlinear systems. Furthermore for long-range intermolecular interactions, Mvogo $et.\,al$ \cite{Mvogo2013,Mvogo2014} showed the existence of elliptic-type solutions in $\alpha$-helix protein chain when taking into account inter spine coupling, while Nji $et.\,al$ \cite{Nji2018} equally investigated on the propagation of periodic soliton wavetrains which is characterised by the $dn$ function, when dealing with a longitudinal interacting amide-I exciton chains of $\alpha$-helix proteins. These previous studies on the propagation of elliptic solitons in various nonlinear media among many others, serves like a big motivation behind the current investigation. In fact, this study seeks to elucidate on the dynamics of elliptic solitons in $\alpha$-helix protein chain, experiencing both exciton-exciton and exciton-phonon  nonlinear coupling forces.

In sec. \textbf{II}, we consider a model Hamiltonian of $\alpha$-helix protein chain which is inherently described by spins. The Hamiltonian is eventually transformed to the classical Davydov Hamiltonian which exclusively deals with boson operators, with the aid of the Dayson-Maleev transformation. Such transformation gives the perfect opportunity to incorporate the nonlinear exciton-exciton coupling. By exploring the quantum and classical mechanical approaches, we basically derived the original equations obtained by Davydov, with the addition of the nonlinear force term introduced by the exciton-exciton interactions. We carry out the adiabatic approximation in sec. \textbf{III} to obtain the discrete nonlinear Schr$\ddot{o}$dinger amplitude equation. It is shown that the exciton-exciton interactions enhance energy localization in the chain via the process of modulational instability. The various regions of the observation of localized modes in the $(q,\,Q)$ plane are equally identified. We numerically investigate the evolution of plane waves in the system to complement the predictions from the modulational instability analysis. Analytic solution of the discrete amplitude equation are equally obtained via the Jacobian elliptic method. We further explore the continuum limit approximation in sec. \textbf{IV} to obtain the inhomogeneous nonlinear Schr$\ddot{o}$dinger equation; with the inhomogeneity induced by the exciton-exciton coupling. In the absence of the exciton-exciton coupling, we observe periodic trains of excitons that signifies how energy of the ATP hydrolysis is uniformly distributed over the protein chain. However when the soliton velocity exceeds that of the phonon, the system supports the propagation of dark solitons which is characterized by the $sn$ function. Solutions of the inhomogeneous nonlinear Schr$\ddot{o}$dinger equation generally depicts periodic exciton modes, which is characterized by the elliptic $cn$ function. We numerically solve the discrete and coupled equations, in order to investigate on the evolution of the elliptic solitons in the protein chain. We finally conclude in sec. \textbf{V} by highlighting the important results obtained, and articulating on some brighter perspectives.   
\section{model equations of the coupled system}
As a model, we consider a hydrogen-bonded chain of $\alpha$-helix protein made up of N-peptide groups which is characterised by two levels. The amide-I exciton is completely absent in the ground state and created in the excited state; which eventually spreads to neighbouring molecules with the aid of the hopping parameter $J$. The Hamiltonian of each two-level molecule described by a spin one-half operator $s_{n}$ is given by \cite{Dicke1954,Xiao1998}
\begin{eqnarray}\label{Alain01}
H=\hbar\omega_{0}\sum_{n}s_{n}^{z}&&-\frac{1}{2}J\sum_{n,\delta}(s_{n}^{+}s_{n+\delta}^{-}+s_{n}^{-}s_{n+\delta}^{+})\nonumber\\
-D\sum_{n,\delta}(s_{n}^{z}+&&\frac{1}{2})(s_{n+\delta}^{z}+\frac{1}{2})\nonumber\\
+\chi\sum_{n}(U_{n+1}-&&U_{n})(s_{n}^{z}+\frac{1}{2})+H_{ph},
\end{eqnarray}
where
\begin{equation}\label{Alain02}
H_{ph}=\sum_{n}\Big(\frac{P_{n}^{2}}{2M}+\frac{1}{2}K(U_{n+1}-U_{n})^{2}\Big).
\end{equation}
In Eq. (\ref{Alain01}), the operators $s_{n}^{+}$ and $s_{n}^{-}$ indicate excitation and
de-excitation, respectively, between two levels with an excitation energy $\hbar\omega_{0}$ at the $n$th site, $\delta$ runs over the nearest neighbors of $n$, and
\begin{equation}\label{Alain03}
[s_{i}^{z},s_{j}^{\pm}]=\pm s_{i}^{\pm}\delta_{ij},\,\,\,\,[s_{i}^{+},s_{j}^{-}]=2s_{i}^{z}\delta_{ij}.
\end{equation}
Concretely in Eq. (\ref{Alain01}), the first term corresponds to the energy of the non-interacting molecules, the second term describes propagation of the excitation associated with the transition dipole moment $d$ and the third term comes from electrostatic interaction between two amide-I excitons that incorporates the static dipole moment $\mu$. The fourth term is the exciton phonon interaction Hamiltonian, with $\chi$ being the exciton-phonon coupling constant which is generally obtained from the temperature dependence of the vibrational  $C$=$O$ modes, observed in infrared spectroscopy experiments. Moreover, the constants $J=2d^{2}/4\pi\epsilon R^{3}$ and $D=2\mu^{2}/4\pi\epsilon R^{3}$, where both the transition and static dipole moments are parallel to the chain direction, $R$ is the nearest neighbor spacing. It should be noted that $J$ and $D$ have different numerical values. $J$ is determined by the transition dipole moment of the amide-I oscillator and the term with $J$ describes the transfer of excitations through the resonance interaction between a molecule in an excited dipole state with an identical non-excited molecule while $D$ is determined by the static electric moment of the amide-I oscillator and describes the electrostatic interactions between two excited molecules, i.e., the electrostatic interaction between two excitons. The value of $J$ can be derived from the experimental data of infrared spectra and it was found that $J=1.55$ x $10^{-22} J$ \cite{Nevskay1976}. A rough estimate of the static dipole moment $\mu$ can be made from the charge distribution of the static $C$=$O$ bond, which is of the same order of magnitude as $d$. Practically, the value of $D$ is difficult to calculate because of the effects of polar $N$-$H$ bonds and solvent, hence we always assume that $D\approx J$.

For the phonon Hamiltonian $H_{ph}$ given in Eq. (\ref{Alain02}), $P_{n}$ and $u_{n}$ respectively denote the momentum and
position operators for longitudinal displacement of the peptide group of site $n$. $M$ is the mass of the peptide group and $K$ is the spring constant of the hydrogen bond.

The Hamiltonian (\ref{Alain01}) exclusively deals with spins, explaining why it is in terms of Pauli operators. This is contrary to the classical Davydov Hamiltonian that deals with boson operators. However, the spin operators in Eq. (\ref{Alain01}) can be transformed to the Davydov Hamiltonian with the aid of Dyson-Maleev transformation and boson commutation relations to obtain \cite{Xiao1998}  
\begin{eqnarray}\label{Alain04}
H=-\frac{1}{2}N\omega_{0}+\omega_{0}\sum_{n}a_{n}^{+}a_{n}&&-\frac{1}{2}J\sum_{n,\delta}(a_{n}^{+}a_{n+\delta}+a_{n}a_{n+\delta}^{+})\nonumber\\
-D\sum_{n,\delta}a_{n}^{+}a_{n}a_{n+\delta}^{+}a_{n+\delta}+\chi\sum_{n}&&(U_{n+1}-U_{n})a_{n}^{+}a_{n}+H_{ph},
\end{eqnarray}
where the term with four boson operators is omitted, because it describes the simultaneous excitations of two amide-I excitons at one molecule. 

Since we are interested with multi-exciton states, it is therefore natural to consider excitons as coherent states whose basis states can be represented in terms of Glauber coherent states \cite{Glauber1958}
\begin{equation}\label{Alain05}
|\Psi\big> =|\left\lbrace \alpha_{n}\right\rbrace \big>   =\prod_{n}|\alpha_{n}\big>,
\end{equation}
as the ansatz for the eigenstates of $H$. The coherent states satisfy the following equation,
\begin{equation}\label{Alain06}
a_{n}|\alpha_{n}\big>   = \alpha_{n}|\alpha_{n}\big>,
\end{equation}
where $\alpha_{n}$ is a complex eigenvalue. Taking $\big<\Psi|H|\Psi\big>$ as the usual classical Hamiltonian function, the time evolution of $\alpha_{n}$ proceeds according to the time-dependent Schr$\ddot{o}$dinger equation
\begin{eqnarray}\label{Alain07}
i\hbar\frac{\partial\alpha_{n}}{\partial t} = \hbar\omega_{0}\alpha_{n}-J(\alpha_{n+1}+\alpha_{n-1}&&)\nonumber\\
-D(|\alpha_{n+1}|^{2}+|\alpha_{n-1}|^{2})\alpha_{n}+\chi(U_{n+1}&&-U_{n})\alpha_{n}.
\end{eqnarray}
While the dynamics of the classical lattice coordinates $U_{n}$, is governed by
\begin{equation}\label{Alain08}
M\frac{\partial^{2}U_{n}}{\partial t^{2}}= K(U_{n+1}+U_{n-1}-2U_{n})+ \chi(|\alpha_{n}|^{2}-|\alpha_{n-1}|^{2}).
\end{equation}
By using the transformation $\alpha_{n}=\phi_{n}exp[-i/\hbar(\hbar\omega_{0}-2J)t]$, Eqs. (\ref{Alain07}) and (\ref{Alain08}) can be rewritten as
\begin{subequations}\label{Alain09}
\begin{align}
i\hbar\frac{\partial\phi_{n}}{\partial t} =&-J(\phi_{n+1}+\phi_{n-1}-2\phi_{n})-D(|\phi_{n+1}|^{2}+\nonumber\\
|\phi_{n-1}|^{2})\phi_{n}&+\chi(U_{n+1}-U_{n})\phi_{n},\\
M\frac{\partial^{2}U_{n}}{\partial t^{2}}= K(&U_{n+1}+U_{n-1}-2U_{n})+ \chi(|\phi_{n}|^{2}-|\phi_{n-1}|^{2}).
\end{align}
\end{subequations}

In the absence of the exciton-exciton coupling (i.e. $D=0$), Eq. (\ref{Alain09}) models the original equation obtained by Davydov, which describe the time evolution of amide-I vibrational energy coupled to displacements of the hydrogen-bonded chain of peptide groups. Furthermore, the distribution of the amide-I energy over the individual peptide groups of the chain is determined by the quantity $|\phi_{n}(t)|^{2}$.

\section{Modulational instability and discrete periodic excitons in the adiabatic limit}
Basically, the principle behind adiabatic approximation is to minimise the delay necessary for the entire lattice to readjust to perturbations induced by a  $C$=$O$ bond excitations. Consequently in this adiabatic approximation (i.e. $\dot{u}_{n}=0$), Eq. (\ref{Alain09}b) becomes
\begin{equation}\label{Alain10}
U_{n+1}-U_{n}= -(\frac{\chi}{K})|\phi_{n}|^{2},
\end{equation}
reducing Eq. (\ref{Alain09}a) to 
\begin{eqnarray}\label{Alain11}
i\frac{\partial\phi_{n}}{\partial t} +\beta(\phi_{n+1}+\phi_{n-1}&&-2\phi_{n})+\gamma_{1}(|\phi_{n+1}|^{2}+|\phi_{n-1}|^{2})\phi_{n}\nonumber\\
+\,\,\,\,\,\gamma_{2}|\phi_{n}|^{2}\phi_{n}=0&&,
\end{eqnarray}
where $\beta=J/\hbar,\,\gamma_{1}=D/\hbar,\,\gamma_{2}=\chi^{2}/\hbar K$. Equation (\ref{Alain11}) is the discrete non-linear Schr$\ddot{o}$dinger (DNLS) amplitude equation in which both the exciton-exciton and exciton-phonon coupling constants provides the necessary non-linear force terms in the adiabatic limit. 

We now investigate the modulational instability (MI) analysis of the discrete amplitude equation (\ref{Alain11}), in order to establish the necessary conditions for energy localization in the $\alpha$-helix protein chain. It is important to note that Eq. (\ref{Alain11}) has exact plane wave solutions in the form $\phi_{n}(t)=\phi_{0}\,exp[i(qn-\omega t)]$, provided the relation between the wavenumber $q$, the angular frequency $\omega$, and the amplitude $\phi_{0}$ is given as
\begin{equation}\label{Alex1}
\omega=4\beta\,sin^{2}\left(\frac{q}{2}\right)-\Big[2\gamma_{1}+\gamma_{2}\Big]|\phi_{0}|^2.
\end{equation}

The MI of amplitude and phase of this plane wave solution is achieved by introducing small perturbations $\theta_{n}(t),\,\,\theta'_{n}(t)$, to the amplitude and phase respectively, so that one can linearize the equation of the envelope of the carrier wave. Therefore we look for a solution of the form $\phi_{n}(t)=\Big[\phi_{0}+\theta_{n}(t)\Big]\,exp[i(qn-\omega t+\theta'_{n}(t))]$, where $\theta_{n}(t),\,\,\theta'_{n}(t)$ are assumed to be small perturbation parameters of the system with frequency $\Omega$ and wave number $Q$. Introducing this perturbation solution in (\ref{Alain11}), using (\ref{Alex1}) and after linearization with respect to constant amplitudes, one obtains the following dispersion relation
$$\Big[\Omega-2\beta\,sin(Q)sin(q)\Big]^{2}=\Big[4\beta\,sin^{2}(Q/2)cos(q)\Big]\Big[4\beta\,sin^{2}(Q/2)$$
\begin{equation}\label{Alex2}
\times cos(q)-2(\gamma_{2}-4\gamma_{1})|\phi_{0}|^2-12\gamma_{1}sin^{2}(Q/2)|\phi_{0}|^{2}\Big].
\end{equation}

Equation (\ref{Alex2}) determines the condition for the stability of a plane wave with wavenumber $q$ in the $\alpha$-helix protein chain. If the RHS of Eq. (\ref{Alex2}) is negative, then the perturbation grows exponentially and MI is possible only if the initial amplitude $|\phi_{0}|$ exceeds the threshold amplitude $|\phi_{0,cr}|$ defined as follows
\begin{equation}\label{Alex3}
|\phi_{0}|^{2}>|\phi_{0,cr}|^{2}=\frac{2\beta\,sin^{2}(Q/2)cos(q)}{(\gamma_{2}-4\gamma_{1})+6\gamma_{1}sin^{2}(Q/2)}, 
\end{equation}
with this instability region constraint to 
\begin{equation}\label{Alex4}
\Big[(\gamma_{2}-4\gamma_{1})+6\gamma_{1}sin^{2}(Q/2)\Big]cos(q)>0. 
\end{equation}
For $\gamma_{1}>0,\,\gamma_{2}>0,\,cos(q)>0$, MI is only observed when the strength of the exciton-phonon coupling far exceeds that of the exciton-exciton coupling (.i.e $\gamma_{2}>4\gamma_{1}$). For $\gamma_{2}=0$, a plane wave will be unstable provided $sin^{2}(Q/2)>2/3$, irrespective of the value of $\gamma_{1}.$ Finally in the absence of exciton-exciton coupling (.i.e $\gamma_{1}=0$), we obtain the exact results by Yuri Kivshar in the study of localized modes in a chain with nonlinear on-site potential \cite{s26}. 

\begin{figure*}
 \centering
 \includegraphics[width=20cm, height=10cm]{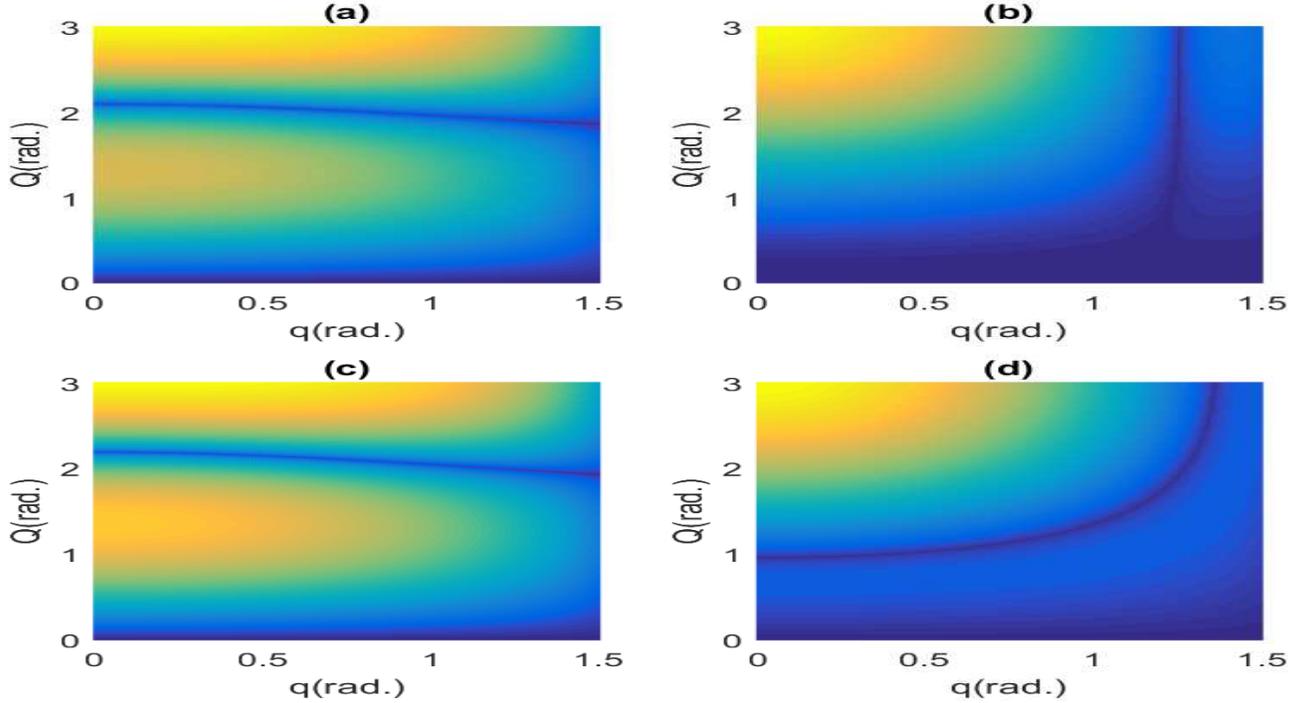}
 \caption{\label{MIR2}(color online) Regions of MI in the $(q,\,Q)$ plane indicated by the yellow area(s) and according to Eq. (\ref{Alex2}). This is for parameters $\beta=1.47$ x $10^{12}s^{-1}$, and $|\phi_{0}|^{2}=1.00$. (a) $\gamma_{1}=3.16$ x $10^{12}s^{-1}$,\,$\gamma_{2}=6.20$ x $10^{11}s^{-1}$, (b) $\gamma_{1}=1.50$ x $10^{11}s^{-1}$,\,$\gamma_{2}=6.20$ x $10^{11}s^{-1}$, (c) $\gamma_{1}=3.16$ x $10^{12}s^{-1}$,\,$\gamma_{2}=0.00$, (d) $\gamma_{1}=0.00$,\,$\gamma_{2}=6.20$ x $10^{11}s^{-1}$.}
 \end{figure*}

Figure \ref{MIR2} shows the profile of the MI conditions on the $(q,\,Q)$ plane for $cos(q)>0$ (i.e. $0\le q<\pi/2$). We maintain the value of J to be $1.55$ x $10^{-22} J$ (.i.e $\beta=1.47$ x $10^{12}s^{-1}$), as derived from the experimental data of infrared spectra \cite{Nevskay1976}, and set $|\phi_{0}|^{2}=1.00$ without loss of generality. The thick blue regions generally depicts areas of plane wave stability, while the yellowish areas signifies regions of unstable plane waves; strongly indicating the existence of discrete solitons. 

In fact, Fig. \ref{MIR2}(a) clearly shows the regions of MI on the $(q,\,Q)$ plane with the protein parameters of the system obtained from Ref. \cite{Scott1992}. We have that the exciton-exciton coupling constant, $D\approx 3.33$ x $10^{-22}J$, the exciton-phonon coupling constant $\chi = 62pN$, and the spring constant $K=58.8N/m$; which effectively leads to the values $\gamma_{1}=3.16$ x $10^{12}s^{-1}$ and $\gamma_{2}=6.20$ x $10^{11}s^{-1}$. We observe that discrete solitons are easily traced in the protein chain for perturbation wave numbers $0.50\le Q\le 1.90$ and $Q\ge 2.30$. An increase in the wave number $q$ in these regions, destroys the solitons by favoring the stability of the plane waves. Figure \ref{MIR2}(b) gives the stability/instability regions in the $(q,\,Q)$ plane for $\gamma_{1}=1.50$ x $10^{11}s^{-1}$, with other parameter values maintained. Note that the decrease in the $\gamma_{1}$ value is temperature dependent, because variations of the orientations of the peptide groups are mainly due to the
rotation around the single bonds connected to the $C_{\alpha}$ atoms. Since we are interested in one degree of freedom, the exciton-exciton coupling constant, $D$, is now inextricably linked to temperature by the relation \cite{Xiao1998,Bergethon1990}
\begin{equation}\label{Alex4}
D(T)=\frac{2}{k_{B}T}\Big(\frac{1}{4\pi\epsilon_{0}}\frac{2\mu^{2}}{R^{3}}\Big)^2, 
\end{equation} 
where $k_{B}$ is the the Boltzmann constant. An increase in temperature clearly constraint the identification of discrete bright solitons in a single region where $Q\ge 1.70$ as in Fig. \ref{MIR2}(b); which is more prominent with small values of wave number $q$. In the absence of exciton-phonon coupling (i.e. $\gamma_{2}=0$), the stability/instability curve is given in Fig. \ref{MIR2}(c); with similar profile to that in Fig. \ref{MIR2}(a). However, the intensity of the yellowish region in Fig. \ref{MIR2}(c) supersedes that in Fig. \ref{MIR2}(a), greatly suggesting that the exciton-phonon coupling favors plane wave stability. Lastly for $\gamma_{1}=0$, we obtain the stability/instability curve in Fig. \ref{MIR2}(d); clearly showing that soliton-like modes in the protein chain can only be observed in a single region. An increase in temperature generally leads to a decrease in $\gamma_{1}$, hence enhancing the existence of Davydov soliton \cite{Xiao1998}, as reflected in Figs. \ref{MIR2}(b,d).

We now perform numerical simulations of the amplitude Eq. (\ref{Alain11}), in order to give a comprehensive account on the evolution of plane waves and energy localization in the $\alpha-$helix protein chain under small perturbations. Consequently, we consider the initial condition 
\begin{equation}\label{Alex12}
\phi_{n}(t=0)=\phi_{0}[1+\Lambda_{0}\,cos(Qn)]cos(qn), 
\end{equation} 
with periodic boundary conditions $\phi_{0}=\phi_{N},\,\phi_{N+1}=\phi_{1}$, where $N=200.$

\begin{figure}[htp]
\centering
\includegraphics[width=10.0cm, height=8.0cm]{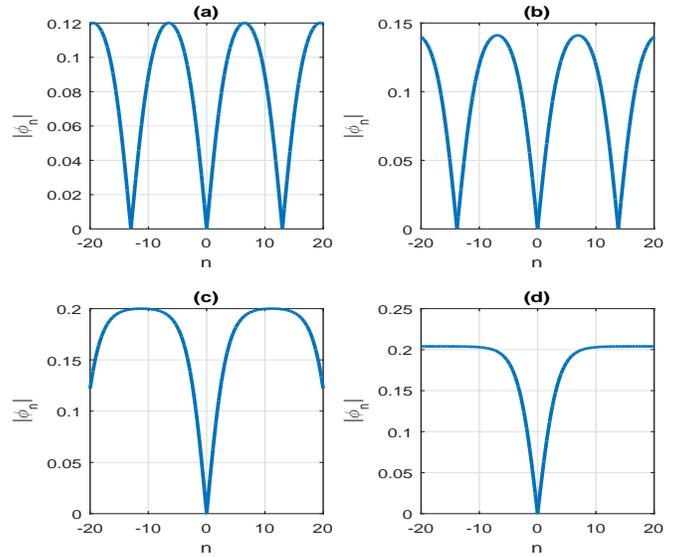}
\caption{\label{sn1}\,(color online) Magnitude of spatial profile of discrete excitons in the $\alpha-$helix protein chain according to solution (\ref{Alex5}). The parameters are given in Eqs. (\ref{Alex11}), with values $p=\pi,\,k=0.3,\,t=0.4,\,\zeta=0.2,\,\nu=0.00,\,\beta=1.47$ x $10^{12}s^{-1}$, $\gamma_{1}=3.16$ x $10^{12}s^{-1}$,\,and $\gamma_{2}=6.20$ x $10^{11}s^{-1}$. This is for $(a)\,m=0.60\,(b)\,m=0.70\,(c)\,m=0.98\,(d)\,m=1.00\,.$}
\end{figure}

The regions of MI shown in Fig. \ref{MIR2} may induce the formation of localized modes. Consequently we now use the Jacobian elliptic function analytical method \cite{Dai2006,Liu2001,Tiofack2007}, to show the existence of this kind of elliptic Davydov solitons in the $\alpha$-helix protein chain. Before applying this analytical method to Eq. (\ref{Alain11}), it is important for us to consider the following transformations 
\begin{equation}\label{Alex5}
\phi_{n}=e^{i\eta_{n}}\rho_{n}(\xi_{n}),\,\eta_{n}=pn+\omega t+\zeta,\,\xi_{n}=kn+ct+\nu. 
\end{equation} 
Upon substitution of the ansatz (\ref{Alex5}) into Eq. (\ref{Alain11}), and, separating the real and imaginary parts yield:
\begin{subequations}\label{Alex6}
\begin{align}
-(\omega+2\beta)\rho_{n}&+\gamma_{1}(\rho_{n+1}^{2}+\rho_{n-1}^{2})\rho_{n}+\gamma_{2}\rho_{n}^{3}\nonumber\\
+\,\,\,\,\,\, \beta(&\rho_{n+1}+\rho_{n-1})cos(p)=0,\\
c\rho'_{n}+ \beta(&\rho_{n+1}-\rho_{n-1})sin(p)=0.
\end{align}
\end{subequations}
We use the following series expansion as a solution of Eqs. (\ref{Alex6}):
\begin{equation}\label{Alex7}
\rho_{n}(\xi_{n})=a_{0}+a_{1}\,sn(\xi_{n}), 
\end{equation} 
where $sn(\xi_{n})\equiv sn(\xi_{n},m)$, and $m$ is the modulus of Jacobian elliptic functions which is constraint by $0\le m\le 1$. The constants $a_{0},\,a_{1}$ are to be determined, and using the identity
\begin{equation}\label{Alex8}
sn(A+B)=\frac{sn(A)cn(B)dn(B)+sn(B)cn(A)dn(A)}{1-m^{2}sn^{2}(A)sn^{2}(B)}, 
\end{equation} 
we obtain  
\begin{equation}\label{Alex9}
\rho_{n+1}(\xi_{n})=a_{0}+a_{1}\Big[\frac{sn(\xi_{n})cn(k)dn(k)+sn(k)cn(\xi_{n})dn(\xi_{n})}{1-m^{2}sn^{2}(\xi_{n})sn^{2}(k)}\Big], 
\end{equation} 
\begin{equation}\label{Alex10}
\rho_{n-1}(\xi_{n})=a_{0}+a_{1}\Big[\frac{sn(\xi_{n})cn(k)dn(k)+sn(k)cn(\xi_{n})dn(\xi_{n})}{1-m^{2}sn^{2}(\xi_{n})sn^{2}(k)}\Big]. 
\end{equation} 
We further substitute the expansions (\ref{Alex7}), (\ref{Alex9}), (\ref{Alex10}) into Eqs. (\ref{Alex6}); clearing the denominator and
setting the coefficients of all powers like $sn^{i}(\xi_{n})\,(i = \,0,\, 1,\, 2,\, 3)$ to zero. After some long calculations, this leads us to the following solutions
$$\omega\,\,=\frac{-4m^{2}\beta\gamma_{1} cos(p)sn^{4}(k)cn(k)dn(k)}{[2\gamma_{1}+\gamma_{2}-2\gamma_{1}(m^{2}+2)sn^{2}(k)+4m^{2}\gamma_{1}sn^{4}(k)]}$$
\begin{subequations}\label{Alex11}
\begin{align}
+&\,\,2\beta[cos(p)cn(k)dn(k)-1],\\
c\,\,\,&=-2\beta sin(p)sn(k),\\
a_{0}&=0,\\
a_{1}^{2}&=\frac{-2m^{2}\beta cos(p)sn^{2}(k)cn(k)dn(k)}{[2\gamma_{1}+\gamma_{2}-2\gamma_{1}(m^{2}+2)sn^{2}(k)+4m^{2}\gamma_{1}sn^{4}(k)]}.
\end{align}
\end{subequations}
The constants $p,\,k,\,\zeta,\,\nu$, are given arbitrary values in order to obtain the most appropriate discrete solution (\ref{Alex5}), as shown in Fig. \ref{sn1}. As the values of the modulus $m$ increases from $0.60$ to $1.00$, the periodic dark excitons in the protein chain degenerates into a single dark soliton in Fig. \ref{sn1}(d).  
\section{Excitons in the coupled system}
\subsection{Analytic solutions in the continuum limit approximation}
In order to effectively comprehend the amide-I $\alpha$-helix protein chain dynamics in the exciton-exciton and exciton-phonon couplings regime, it is incumbent on us to solve the discrete coupled system (\ref{Alain09}). However, it is cumbersome to solve Eqs. (\ref{Alain09}) as such due to its high non-linearity and discreteness. Therefore it is quite useful to explore the continuum limit approximation in the low-temperature and long-wavelength limit. Consequently, we carry out a continuous-limit approximation by introducing the spatial variable $x=\varepsilon na$, where $a$ is the lattice spacing and $\varepsilon << 1$ is a perturbation parameter which physically accounts for the interaction of the system with its quantum environment. This leads to
\begin{subequations}\label{Alain13}
\begin{align}
\phi_{n\pm 1}(t)&= \phi(x,t) \pm \varepsilon a\frac{\partial\phi(x,t)}{\partial x}+\frac{\varepsilon^{2}a^{2}}{2!}\frac{\partial^{2}\phi(x,t)}{\partial x^{2}}+\bigcirc(\varepsilon^{3}),\\
U_{n\pm 1}(t)&=U(x,t) \pm \varepsilon a\frac{\partial U(x,t)}{\partial x}+\frac{\varepsilon^{2}a^{2}}{2!}\frac{\partial^{2}U(x,t)}{\partial x^{2}}+\bigcirc(\varepsilon^{3}).
\end{align}
\end{subequations}
Also in the weak exciton-phonon coupling regime, we perturb $\chi$ to the order $\varepsilon$ and transform Eqs. (\ref{Alain09}) to
\begin{subequations}\label{Alain14a}
\begin{align}
i\hbar\phi_{t}+ J\varepsilon^{2}a^{2}\phi_{xx}+2D|\phi|^{2}\phi =   -\varepsilon^{2}D&\big[(\phi\phi_{x}^{*})_{x}+|\phi_{x}|^{2}\big]\phi\,\,\,\,\,\,\,\,\,\,\,\,\,\,\,\,\,\,\,\,\,\,\,\,\,\,\,\,\,\,&\nonumber\\
+\chi\varepsilon^{2}aU_{x}\phi\,\,\,\,\,\,\,\,\,\,\,\,\,\,\,&\\
MU_{tt}-K\varepsilon^{2}a^{2}U_{xx}= \chi\varepsilon^{2}a|\phi|_{x}^{2}.\,\,\,\,\,\,\,\,\,\,\,\,\,\,\,&
\end{align}
\end{subequations}
We shall seek travelling wave solutions of Eqs. (\ref{Alain14a}) in the form of excitations that propagate along the chain with a velocity $v_{0}$ such that
\begin{equation}\label{Alain15}
U(x,t)=U(x-\varepsilon v_{0}t).
\end{equation}
Inserting Eq. (\ref{Alain15}) in Eq. (\ref{Alain14a}b), we get
\begin{equation}\label{Alain16}
U_{x}=\frac{-\chi|\phi|^{2}}{Ka\big(1-v_{0}^{2}/v_{0s}^{2}\big)},
\end{equation}
where $v_{0s}=\sqrt{\frac{Ka^{2}}{M}}$ is the velocity of sound which is greater than the velocity of excitations in the chain i.e. $v_{0s}>v_{0}$. Substituting Eq. (\ref{Alain16}) into Eq. (\ref{Alain14a}a), we get

\begin{equation}\label{Alain17}
i\phi_{t} +P\phi_{xx}+Q|\phi|^{2}\phi+R\big[(\phi\phi_{x}^{*})_{x}+|\phi_{x}|^{2}\big]\phi=0,
\end{equation}
where the coefficients are given by
\begin{subequations}\label{Alain18}
\begin{align}
P&=\frac{J\varepsilon^{2}a^{2}}{\hbar},\\
Q&=\frac{2D}{\hbar}+\frac{\varepsilon^{2}\chi^{2}}{\hbar K\big(1-v_{0}^{2}/v_{0s}^{2}\big)},\\
R&=\frac{\varepsilon^{2}D}{\hbar}.
\end{align}
\end{subequations}
Equation (\ref{Alain17}) is the inhomogeneous NLS equation, which governs the amide-I $\alpha$-helix protein chain dynamics in the continuous limit approximation, with the inhomogeneity being induced by the exciton-exciton coupling. The form of the envelope soliton solution of Eq. (\ref{Alain17}) is given by
\begin{equation}\label{chap09eq4701}
\phi(x,t)=\phi_{0}\,u(X)\,exp[i(\psi(X)+\omega t)],
\end{equation}
where
$$X=X(x,t)=x-ct.$$
$u$ and $\psi$ are real functions of $X$, while $\phi_{0},\,\omega,\,c$ are real constants with $c$ being the speed while $\omega$ is the frequency. Inserting the ansatz (\ref{chap09eq4701}) into Eq. (\ref{Alain17}) and separating the real and imaginary parts gives
\begin{subequations}\label{Alain19}
\begin{align}
P(u''-u\psi'^{2}) + &\phi_{0}^{2}Qu^{3}+(c\psi'-\omega)u+\phi_{0}^{2}R(2u'^{2}+uu''+\nonumber\\
\psi'^{2}u^{2})u\,\,\,\,\,\,\,=\,\,\,\,\,\,&0,\\
P(2u'\psi'+u\psi'')&-cu'-\phi_{0}^{2}R(\psi''u^{2}+2uu'\psi')u =0.
\end{align}
\end{subequations}
At this stage, we assume that
\begin{equation}\label{Alain20}
\psi'=bu^{2} + \sigma,
\end{equation}
where $b$ and $\sigma$ are real constants. Upon substituting (\ref{Alain20}) into Eq. (\ref{Alain19}) and neglecting terms of order $\bigcirc(u^{4})$ and above yields
\begin{subequations}\label{Alain21}
\begin{align}
P(u''-\sigma^{2}u-2\sigma bu^{3}) + &\phi_{0}^{2}Qu^{3}+c(bu^{2}+\sigma)u-\omega u+\nonumber\\
\phi_{0}^{2}R(2u'^{2}+uu''+\sigma^{2}u^{2})u&=\,\,\,\,\,\,\,\,\,\,0,\\
(2P\sigma-c)u'+(4bP-2\phi_{0}^{2}&R\sigma)u'u^{2} =0.
\end{align}
\end{subequations}
Since $u'\ne 0$, $u'u^{2}\ne 0$  in Eq. (\ref{Alain21}b), we respectively obtain $c=2P\sigma,\,\,\phi_{0}^{2}R=2bP/\sigma$ and transform Eq. (\ref{Alain21}a) to
\begin{equation}\label{Alain22}
u''=\Gamma_{1}u-\Gamma_{2}u^{3}-\Gamma_{3}(u^{2}u')',
\end{equation}
where
\begin{equation}\label{Alain23}
\Gamma_{1}=\frac{(\omega-P\sigma^{2})}{P},\,\Gamma_{2}=\frac{\phi_{0}^{2}}{P}(R\sigma^{2}+Q),\,\Gamma_{3}=\frac{\phi_{0}^{2}R}{P}.
\end{equation}

\begin{figure}[htp]
\centering
\includegraphics[width=10.0cm, height=8.0cm]{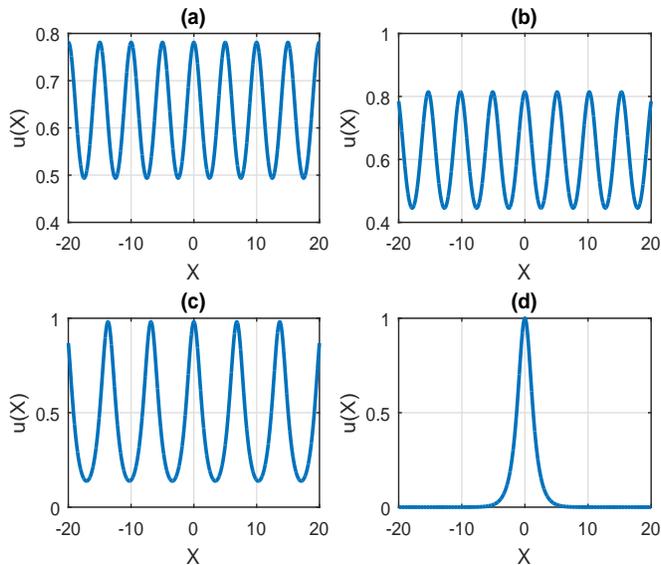}
\caption{\label{cp02dn}\,(color online) Spatial periodic excitons which is solution (\ref{chap09eq4708}). The parameters are
$\phi_{0}=\sqrt{2},\,P=1.00,\,Q=1.00,\,\sigma=1.0,\,\omega=2.00$. This is for
$(a)\,m=0.60\,(b)\,m=0.70\,(c)\,m=0.98\,(d)\,m=1.00\,.$}
\end{figure}

In the absence of the of exciton-exciton coupling (i.e. $D=0,\,\Gamma_{3}=0,\,R=0,\,Q=\varepsilon^{2}\chi^{2}/\hbar K\big(1-\frac{v_{0}^{2}}{v_{0s}^{2}}\big)$, Eq. (\ref{Alain22}) possesses solution of the form \cite{Vyas2006,Hioe1999, Hioe2003}
\begin{equation}\label{chap09eq4708}
u(X)=A\,dn(BX,\,m)+C,
\end{equation}
where $A,\,B$ and $C$ are parameters to be determined, with $m$ being the modulus $(0\leqslant m \leqslant 1)$ of the
elliptic $dn$ function. By substituting (\ref{chap09eq4708}) into Eq. (\ref{Alain22}) and equating coefficients of terms with
the same powers of $dn(BX,\,k)$ to have
\begin{equation}\label{chap09eq4707b}
A=\sqrt{\frac{2(\omega-P\sigma^{2})}{\phi_{0}^{2}Q(2-m^{2})}},\,B=\sqrt{\frac{\omega-P\sigma^{2}}{P(2-m^{2})}},\,C=0.
\end{equation}
Figure \ref{cp02dn} is the $dn$ function which depicts the train of spatial periodic excitons in the amide-I $\alpha$-helix protein chain.
The non localized solutions depicted in Fig. \ref{cp02dn}(a-c) shows that the energy of the ATP hydrolysis is locally distributed over the the $\alpha$-helix protein chain. This energy may be self-localized to form a single bright soliton on one side of protein chain for $m=1.0$ (i.e. it is as a result of superposition of continuous plane waves oscillating at a constant frequency shift), as shown in Fig. \ref{cp02dn}(d). However from experimental standpoint, the observation of such localized nature of energy transfer in a protein chain is very rare. In fact, $m$ may be considered as a pulse dispersion parameter with the periodic excitons mainly conveying coded
information being processed in the amide-I $\alpha$-helix protein chain in the spatial domain.

\begin{figure}[htp]
\centering
\includegraphics[width=9.0cm, height=8.0cm]{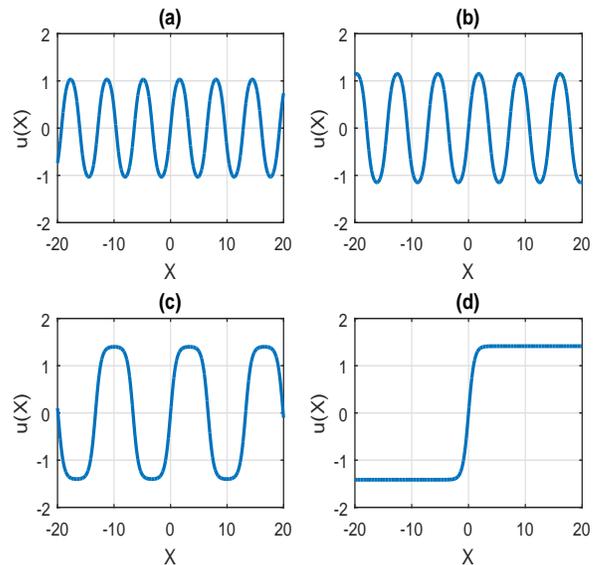}
\caption{\label{Dark1}\,(color online) Profile of Jacobian $sn$ function along the $\alpha$-helix protein chain according to solution (\ref{Alain18b}). The parameters are $\phi_{0}=1.00,\,P=1.00,\,Q=-1.00,\,\sigma=2.00,\,\omega=3.00$. This is for $(a)\,m=0.60\,(b)\,m=0.70,(c)\,m=0.98\,(d)\,m=1.00\,.$}
\end{figure}

The Cherenkov-like effect initially demonstrated by Cherenkov in 1934 \cite{Landau1960}, may equally be identified in the no exciton-exciton (i.e. $D=0$) coupling regime; provided the soliton velocity exceeds that of the phonon (.i.e $v_{0}>v_{0s}$ as in ref \cite{Mesquita1998}). Consequently, we equally obtain dark profile soliton of Eq. (\ref{Alain22}) given by
\begin{equation}\label{Alain18b}
u(X)=\sqrt{\frac{2m^{2}(P\sigma^{2}-\omega)}{\phi_{0}^{2}|Q|(1+m^{2})}}\,sn\Big[\sqrt{\frac{(P\sigma^{2}-\omega)}{P(1+m^{2})}}X,\,m\Big],
\end{equation}
provided the frequency $\omega<P\sigma^{2}$. This leads to the propagation of the elliptic $sn$ exciton in the $\alpha$-helix protein chain as shown in Fig. \ref{Dark1}; with the profile of this periodic excitons degenerating to a kink soliton for $m=1.00$ as shown in Fig. \ref{Dark1}(d). Solution (\ref{Alain18b}) manifest the same structural features as the spatial profile of discrete excitons in the adiabatic limit, as dictated by Eq. (\ref{Alex7}). 

For $\Gamma_{3}\ne 0,$ traveling wave solutions of Eq. (\ref{Alain22}) can be obtained by using the following series expansion \cite{Vyas2006,Hioe1999, Hioe2003}
\begin{equation}\label{Alain24}
u(X)=c_{0}+c_{1}\,cn(X)+c_{2}\,cn^{2}(X),
\end{equation}
where $cn(X)\equiv cn(X,m)$ and $c_{0},\,c_{1},\,c_{2}$ are real constants to be determined. Since our goal is to look for spatial periodic solutions, we substitute our proposed solution (\ref{Alain24}) into Eq. (\ref{Alain22}), and equating the coefficients of equal
powers of $cn(X)$ to have:
\begin{subequations}\label{Alain26}
\begin{align}
c_{0}^{2}&=\frac{2m^{2}-1}{4\Gamma_{2}+(2m^{2}-1)\Gamma_{3}},\\
c_{1}^{2}&=\frac{4\Gamma_{1}\Gamma_{2}+(2m^{2}-1)\big[\Gamma_{1}\Gamma_{3}-\Gamma_{2}\big]}{2(1-m^{2})\big[4\Gamma_{2}\Gamma_{3}+(2m^{2}-1)\Gamma_{3}^{2}\big]},\\
c_{2}^{2}&=0.
\end{align}
\end{subequations}

\begin{figure}[htp]
\centering
\includegraphics[width=9.0cm, height=12.0cm]{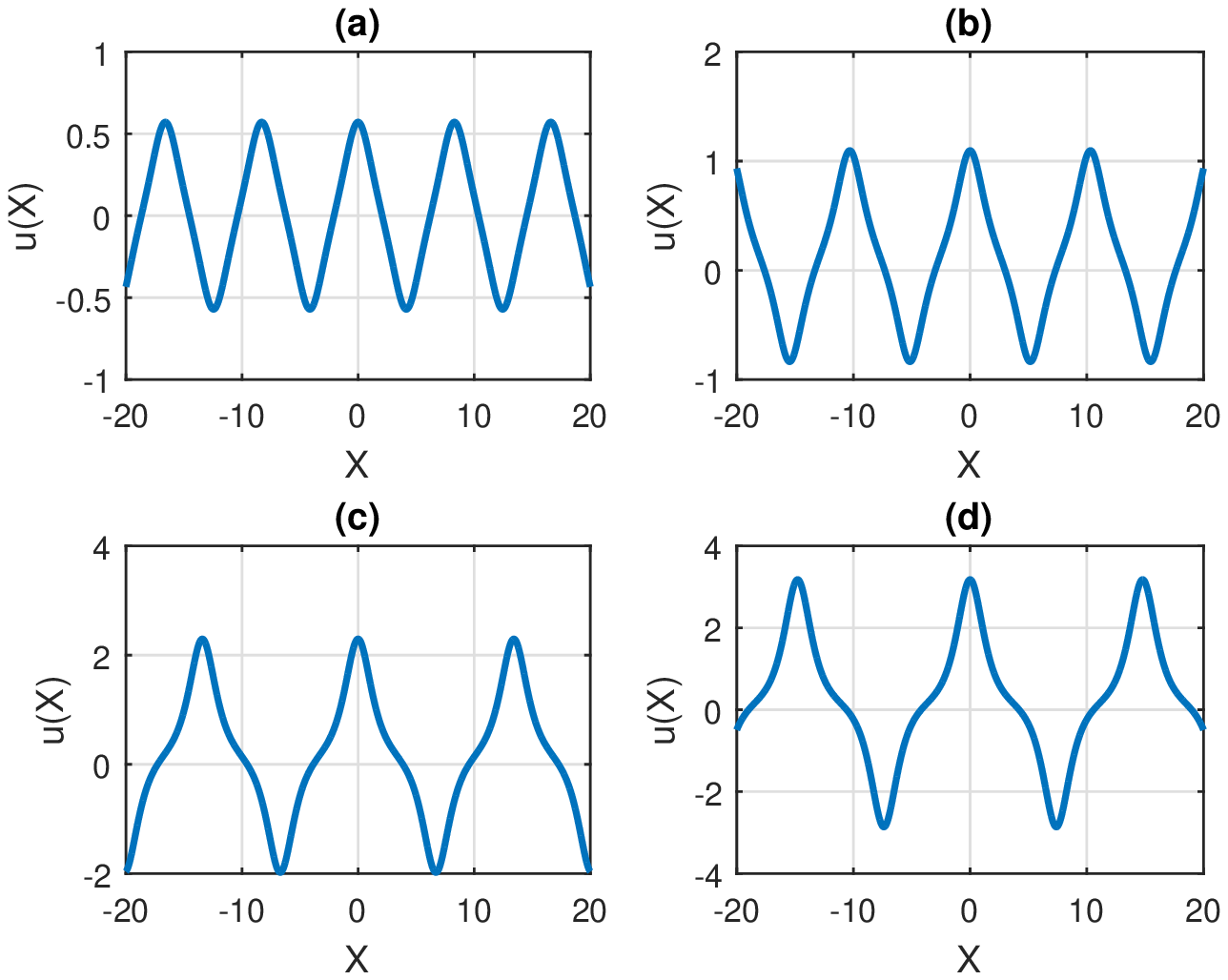}
\caption{\label{Dark}\,(color online) Spatial periodic solitons in the protein chain as governed by solution (\ref{Alain24}). The parameters are $\phi_{0}=\sqrt{2},\,P=1.00,\,Q=1.00,\,\sigma=1.00,\,R=3.00,\,\omega=5.00,\,\Gamma_{1}=4.00,\,\Gamma_{2}=8.00,\,\Gamma_{3}=6.00$. This is for $(a)\,m=0.70\,(b)\,m=0.90,(c)\,m=0.98\,(d)\,m=0.99\,.$}
\end{figure}

\begin{figure}[htp]
\centering
\includegraphics[width=9.0cm, height=12.0cm]{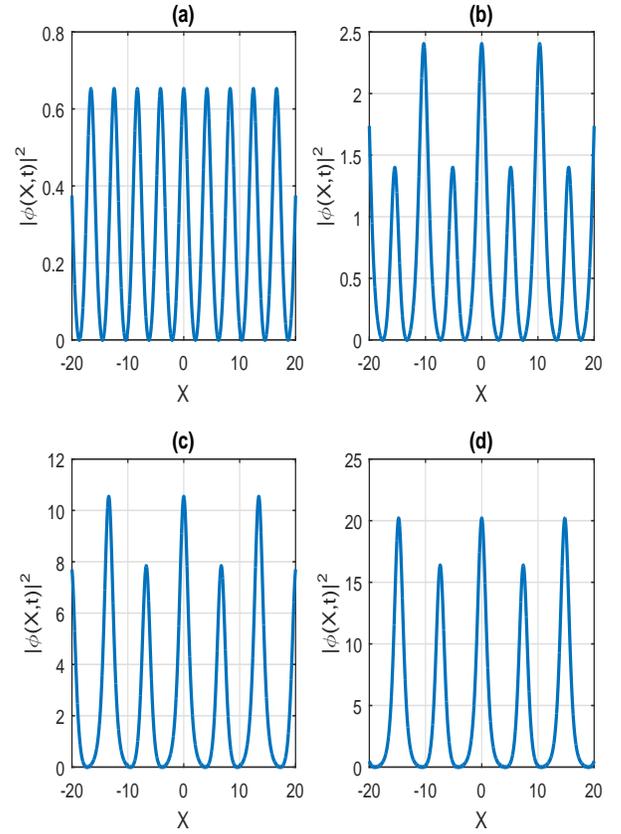}
\caption{\label{SecondQ}\,(color online) Intensity of spatial periodic solitons in the protein chain as governed by solution (\ref{chap09eq4701}). The parameters are $\phi_{0}=\sqrt{2},\,P=1.00,\,Q=1.00,\,\sigma=1.00,\,R=3.00,\,\omega=5.00,\,\Gamma_{1}=4.00,\,\Gamma_{2}=8.00,\,\Gamma_{3}=6.00$. This is for $(a)\,m=0.70\,(b)\,m=0.90,(c)\,m=0.98\,(d)\,m=0.99\,.$}
\end{figure}

Figure (\ref{Dark}) shows the spatial profile of elliptic solitons in the $\alpha-$helix protein chain as the modulus $m$ is varied. On the other hand, the intensity profile of this spatial periodic soliton is given in Fig. (\ref{SecondQ}). 

\begin{figure}[htp]
\centering
\includegraphics[width=9.0cm, height=12.0cm]{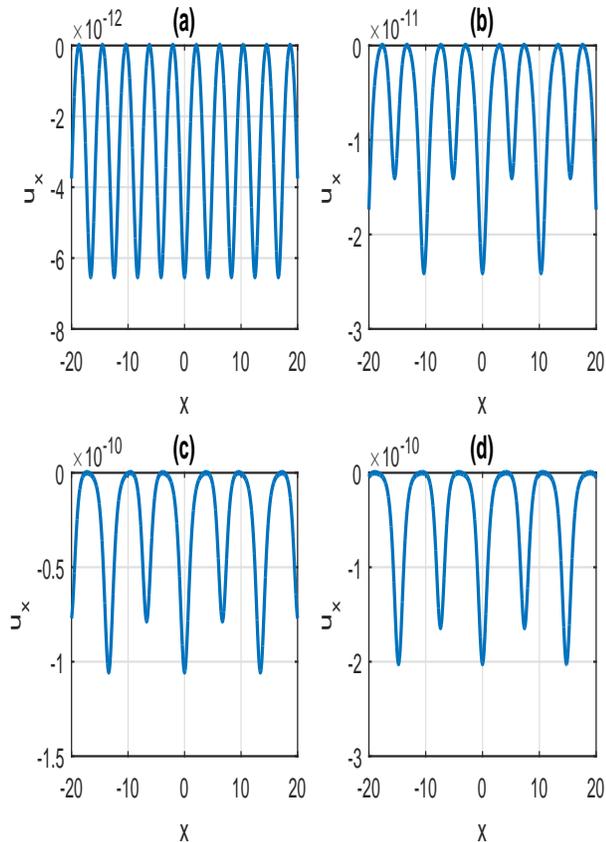}
\caption{\label{ThirdQ}\,(color online) Spatial profile of the displacement of the peptide groups given by solution (\ref{Alain16}). The parameters are $\phi_{0}=\sqrt{2},\,P=1.00,\,Q=1.00,\,\sigma=1.00,\,R=3.00,\,\omega=5.00,\,\Gamma_{1}=4.00,\,\Gamma_{2}=8.00,\,\Gamma_{3}=6.00,\,a=1.00,\,v_{0}^{2}=0.90v_{0s}^{2},\,K=58.8Nm^{-1},$ and the exciton-phonon coupling constant $\chi = 62pN.$ This is for $(a)\,m=0.70\,(b)\,m=0.90,(c)\,m=0.98\,(d)\,m=0.99\,.$}
\end{figure}

We equally plot the spatial profile of the displacement of the peptide groups given by solution (\ref{Alain16}) in Fig. (\ref{ThirdQ}). 
\subsection{Numerical analysis}
A comprehensive study of Eqs. (\ref{Alain09}) is quite cumbersome, because of the high non-linearity and discreteness that characterizes such an equation. In order to effectively comprehend the amide-I $\alpha$-helix protein chain dynamics in the exciton-exciton and exciton-phonon couplings regime, it is incumbent on us to numerically solve the discrete coupled system (\ref{Alain09}). We now consider the following approximation $U_{n+1}-U_{n}\approx \varepsilon a\partial U/\partial x$, in order to obtain the following approximate initial conditions
\begin{subequations}\label{Alex20}
\begin{align}
\phi_{n}(t=0)&\approx\phi_{0}u_{n}\,exp\Big[-i\varepsilon a\big[bu_{n}^{2}+\sigma\big]\Big],\\
U_{n}(t=0)&\approx\frac{\varepsilon\chi|\phi_{0}|^{2}|u_{n}|^{2}}{K\big(1-v_{0}^{2}/v_{0s}^{2}\big)},
\end{align}
\end{subequations}
for the numerical integration of Eqs. (\ref{Alain09}). Without loss of generality, we set $\varepsilon=a=1$ and consider the following three initial conditions:
\begin{enumerate}
\item For $D=0$, the $dn$ soliton solution (\ref{chap09eq4708}) gives:
\begin{equation}\label{Alex21}
u_{n}=\sqrt{\frac{2(\omega-P\sigma^{2})}{\phi_{0}^{2}Q(2-m^{2})}}\,dn\Big[\sqrt{\frac{\omega-P\sigma^{2}}{P(2-m^{2})}}n,\,m\Big].
\end{equation}

\item For $D=0$, the $sn$ soliton solution (\ref{Alain18b}) gives:

\begin{equation}\label{Alex22}
u_{n}=\sqrt{\frac{2m^{2}(P\sigma^{2}-\omega)}{\phi_{0}^{2}|Q|(1+m^{2})}}\,sn\Big[\sqrt{\frac{(P\sigma^{2}-\omega)}{P(1+m^{2})}}n,\,m\Big],
\end{equation}
provided the frequency $\omega<P\sigma^{2}$.

\item For $D\ne 0$, the $cn$ soliton solution (\ref{Alain24}) gives:
$$u_{n}=\sqrt{\frac{2m^{2}-1}{4\Gamma_{2}+(2m^{2}-1)\Gamma_{3}}} \,\,\,\,\,+ $$
\begin{equation}\label{Alex23}
\sqrt{\frac{4\Gamma_{1}\Gamma_{2}+(2m^{2}-1)\big[\Gamma_{1}\Gamma_{3}-\Gamma_{2}\big]}{2(1-m^{2})\big[4\Gamma_{2}\Gamma_{3}+(2m^{2}-1)\Gamma_{3}^{2}\big]}}cn(n,m).
\end{equation}

\end{enumerate}
\section{Discussion and Conclusion}
It is clearly an indisputable fact that proteins are the principal organs used in the generation of energy for the metabolic activities of the cell and by extension to maintain life activities. Consequently there is a growing interest in the field of bioenergetics, because of the quest to fully understand the mechanism of generation and transfer of energy in protein chains. Davydov argued that the energy released by ATP hydrolysis is stored in the amide-I vibration through nonlinear interactions, that traps the energy in the form of a robust entity called solitons. The soliton concept is a crucial idea that should be rigorously considered by biophysicists. Even though it can not explain every aspect of protein chain dynamics, it however serve like a guide for future experimental investigations. In this study, we considered a model Hamiltonian of $\alpha$-helix protein chain which deals with spins, and later transformed it to the classical Davydov Hamiltonian to exclusively handle boson operators. The quantum mechanical approach was then explored to treat the amide-I excitations, while the displacements of longitudinal sound wave along the hydrogen-bonded chain was treated classically. In the adiabatic limit approximation, we have demonstrated that the mechanism of the formation of the localized modes in the protein chain is inextricably linked to the modulational instability phenomena. We further explored all the nonlinearity in the protein chain, in order to enhance the self-trapping mechanism of the amide-I vibrational energy. Such mechanism has been shown to provide a possible solution to the problem of the thermal stability of the Davydov soliton \cite{Xiao1998}. By applying the continuum limit approximation to our coupled system, the inhomogeneous NLS equation was obtained. It was generally observed that the inhomogeneity induced by the exciton-exciton coupling, favors energy localization in the chain as elaborated by the results of numerical simulations. 

In perspective; firstly we intend to use the soliton concept of $\alpha$-helix protein chain to explain a possible mechanism of general anaesthesia. This will be based on the assumption that the binding of an anaesthetic molecule to a protein interferes with soliton propagation. To validate this idea from a theoretical standpoint, we need to calculate the effect of anaesthetic binding on soliton propagation. Secondly, the interaction of a system with its environment is given by the dissipative effect in quantum system, which has attracted many interests in the last decades. Such dissipative effects can be effectively investigated by
adding new term the inhomogeneous NLS equation. A remarkable result of this present study is that the exciton-exciton coupling enhances self-trapping, which accounts for many aspects of protein dynamics.
\section*{Ackowledgments}
N. Oma Nfor appreciates the enriching discussions with members of the \textbf{LaRAMaNS} research group.

\end{document}